\documentclass[twocolumn,prb, showpacs]{revtex4}
\usepackage{graphicx}
\usepackage{dcolumn}
\usepackage{bm}

\begin{document}


\title{Magnetic susceptibility study of hydrated and non-hydrated\\ Na$_x$CoO$_2$$\cdot$yH$_2$O single
crystals}

\author{F. C. Chou$^1$}
\author{J. H. Cho$^{1,2,}$}
\altaffiliation[On leave ]{from Physics Department, Pusan National University, Korea.}
\author{Y. S. Lee$^{1,2}$}
\affiliation{%
$^1$Center for Materials Science and Engineering,
Massachusetts Institute of Technology \\}%
\affiliation{%
$^2$Department of Physics,
Massachusetts Institute of Technology}%

\date{\today}

\begin{abstract}
We have measured the magnetic susceptibility of single crystal
samples of non-hydrated Na$_x$CoO$_2$ ($x \simeq 0.75, 0.67, 0.5,
0.3$) and hydrated Na$_{0.3}$CoO$_2$ $\cdot y$H$_2$O ($y \simeq 0,
0.6, 1.3$).  Our measurements reveal considerable anisotropy
between the susceptibilities with $H\parallel c$ and $H\parallel
ab$. The derived anisotropic $g$-factor ratio ($g_{ab}/g_{c}$)
decreases significantly as the composition is changed from the
Curie-Weiss metal with $x = 0.75$ to the paramagnetic metal with
$x = 0.3$.  Fully hydrated Na$_{0.3}$CoO$_2$$\cdot$1.3H$_2$O
samples have a larger susceptibility than non-hydrated
Na$_{0.3}$CoO$_2$ samples, as well as a higher degree of
anisotropy.  In addition, the fully hydrated compound contains a
small additional fraction of anisotropic localized spins.
\end{abstract}

\maketitle

\section{\label{sec:level1}Introduction\protect\\ }

The discovery of superconductivity below $T \sim 4.5$~K in
Na$_{0.3}$CoO$_2$$\cdot$1.4H$_2$O has engendered much interest in
the family of Na$_x$CoO$_2$$\cdot y$H$_2$O
compounds.\cite{Takada2003}  The non-hydrated compound
Na$_{x}$CoO$_2$ with $x\simeq 0.65-0.75$ has an anomalously large
thermoelectric power.~\cite{Terasaki1997}  Measurements in applied
magnetic fields indicate that spin entropy plays an important role
in the enhanced thermopower in Na$_{0.68}$CoO$_2$.~\cite{Wang2003}
Further work has revealed that the Na$_x$CoO$_2$ material crosses
over from a Curie-Weiss metal ($x>0.5$) to a paramagnetic metal
($x<0.5$), separated by a charge-ordered insulator at
$x=0.5$.~\cite{Foo2004} Based on density-functional calculations,
weak itinerant ferromagnetic fluctuations have been suggested to
compete with weak antiferromagnetism for $x = 0.3$ to $x =
0.7$.~\cite{Singh2003}  Much recent research has focused on
understanding the mechanism for superconductivity in the new
hydrated superconductor. Further progress can be made by examining
the bulk properties of Na$_x$CoO$_2$$\cdot$yH$_2$O as a function
of $x$ and $y$ in order to elucidate the intriguing physics found
on the phase diagram.

In this paper, we present a systematic study of the magnetic
susceptibility of Na$_x$CoO$_2$$\cdot y$H$_2$O (with $x \simeq
0.75, 0.67, 0.5, 0.3$ and $y \simeq 0, 0.6, 1.3$) using
single-crystal samples produced by an electrochemical
de-intercalation method.~\cite{Chou2004,Chou-EC} Our results show
that the susceptibilities of our crystals are consistent with the
behavior reported in powder samples prepared differently via
chemical Br$_{2}$ de-intercalation.  In addition, our studies
reveal that the susceptibility is clearly anisotropic. We further
report how the susceptibility changes as a function of both Na and
H$_2$0 content.  The paper is arranged as follows: Section II
contains experimental details and initial sample characterization.
Section III contains the results of our susceptibility
measurements, along with analysis of our data.  A discussion of
our results and conclusions are in Section IV.

\section{\label{sec:level1}Experimental\protect\\}

Single crystals of Na$_{0.75}$CoO$_2$ were grown using the
floating-zone technique.  After an additional electrochemical
de-intercalation procedure, samples were produced with the final
Na concentrations of $x = 0.75, 0.67, 0.5$ and 0.3, as confirmed
by Electron Microprobe Analysis.  Details of the crystal growth
process, electrochemical de-intercalation, and characterization of
the resulting samples are discussed in depth in Ref. [7].  Powder
neutron diffraction on a crushed single crystal of
Na$_{0.75}$CoO$_2$ grown under similar conditions as the samples
presented here indicates that the crystalline boule consists of a
single structural phase.~\cite{Qing-0.75} A crystal of
Na$_{0.7}$CoO$_2$ was grown using the flux method.  The crystal
was obtained from a melt prepared from powder mixtures of
Na$_{0.75}$CoO$_{2}$+4$\:$NaCl+4$\:$Na$_{2}$CO$_{3}$+B$_{2}$O$_{3}$
which was slowly cooled from 920$^\circ$C to 820$^\circ$C at a
rate of -1$^\circ$C/hr.

A fully hydrated Na$_{0.3}$CoO$_2$$\cdot$1.3H$_2$O crystal was
prepared by enclosing a non-hydrated Na$_{0.3}$CoO$_2$ crystal
within a water vapor saturated environment for $\sim$4 months.
After this period of hydration, the crystal consisted of a single
phase with a $c$-axis lattice constant of  $19.56(8)$~\AA~and with
a superconducting transition temperature near $\sim4.2$~K.
Partially hydrated Na$_{0.3}$CoO$_{2}\cdot 0.6\:$H$_2$O samples
and a fully dehydrated Na$_{0.3}$CoO$_2$ samples were obtained by
driving water out by annealing at temperatures of 120$^{\circ}$C
and 220$^{\circ}$C, respectively, for 15 hours.  Except for a
small fraction of a Co$_3$O$_4$ impurity phase generated in the
fully dehydrated crystal, the partially hydrated and fully
dehydrated phases have c-axis lattice constants of
$13.74(3)$~\AA~and $11.11(5)$~\AA, respectively, which are in
agreement with previously reported values of
Na$_{0.3}$CoO$_2$$\cdot 0.6\:$H$_2$O and
Na$_{0.3}$CoO$_2$.~\cite{Foo2003}  In principal, the fully
dehydrated (FD) phase should be identical to the original
non-hydrated phase.  Powder x-ray diffraction results confirm that
the two phases share the same primary powder peaks; however, the
FD sample contains a somewhat higher degree of defects which we
will discuss further below.

Measurements of the magnetic susceptibility were performed using a
Quantum Design MPMS-XL SQUID magnetometer.  The typical size of
the crystals used in these measurements were $4\times 3\times
1$~mm$^3$ with mass $\sim 50$~mg.  These pieces were easily
cleaved from the larger floating-zone boule with the larger
surface area corresponding to the $ab$-plane. All data were
measured under a magnetic field of 1 Tesla through both zero
cooled and zero-field cooled sequences. Slightly hysteretic
behavior was observed for the $x = 0.75$ sample near temperatures
of $\sim$22 K and $\sim$320 K, consistent with previous
observations.~\cite{Prakhakaran2003,Sales2004} Since the focus of
this paper is not on the hysteretic behavior, all of the data
presented are based on the zero-field cooled results.  A
background correction to the susceptibility has been made on the
crystals with $x=0.67$ and 0.3.  The source of this background is
likely due to a small amount of CoO impurities (about a 7 \%\ mass
fraction) imbedded between the grain boundaries of the particular
starting boule of Na$_{0.75}$CoO$_2$. The CoO correction was
implemented as to minimize the contribution from the
antiferromagnetic transition (T$_{N}$$\sim$290K) of CoO, while
maintaining the same magnitude of the susceptibility as a powder
sample of the same stoichiometry.~\cite{Chou-EC}  No correction
was applied to the Na$_{0.75}$CoO$_2$ and Na$_{0.5}$CoO$_2$
crystals, where no anomaly at T$\sim$290 K is found and the powder
averaged data agrees with the data from stoichiometric
polycrystalline samples. We note, however, there is a broad
maximum near $\sim275$~K for the Na$_{0.67}$CoO$_2$ crystal, which
may be intrinsic, as this is also reported by
others.~\cite{Prakhakaran2003}

\section{\label{sec:level1}Results and Analysis\protect\\ }

\subsubsection{\label{sec:level2}Magnetic susceptibility of non-hydrated
Na$_x$CoO$_2$}

\begin{figure}
\includegraphics[width=3.5in]{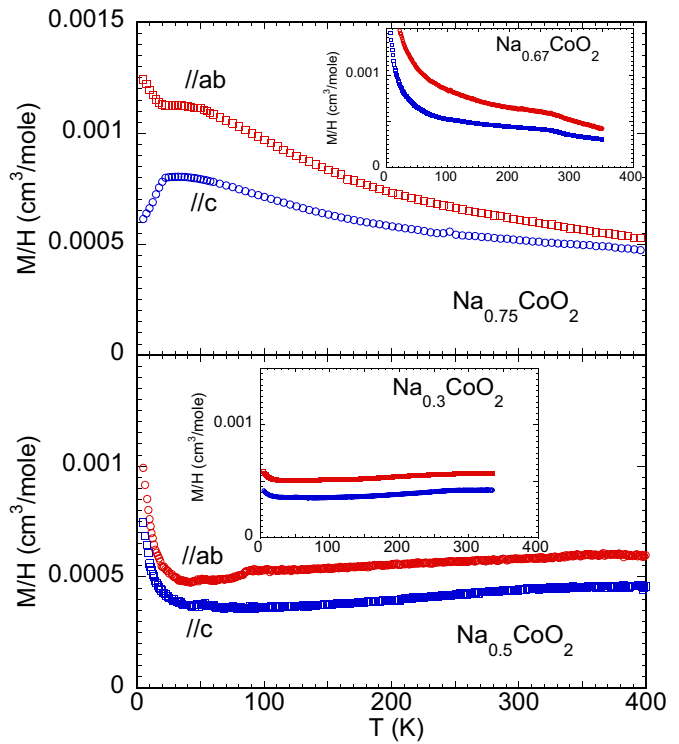}
\caption{\label{fig:eps1} (color online) Magnetic susceptibilities
of Na$_{x}$CoO$_{2}$ (x = 0.75, 0.67, 0.5, and 0.3) under a
magnetic field of 1 Tesla.  The red and blue symbols are for the
magnetic field applied along the ab- and c- directions,
respectively.}
\end{figure}

Our magnetic susceptibility data for Na$_x$CoO$_2$ ($x = 0.3, 0.5,
0.67,$ and 0.75) are shown in Fig.~\ref{fig:eps1}, where a
magnetic field of 1 Tesla was applied parallel to the $ab$- and
$c$- directions.  Curie-Weiss-like behavior is observed for $x >
0.5$, but samples with $x < 0.5$ show a monotonic increase of
$\chi$ with increasing temperature.  Our results, when
powder-averaged, agree with the published powder measurements on
samples prepared using chemical Br$_{2}$
de-intercalation.~\cite{Foo2004} For example, the sample with
$x=0.5$ shows anomalies in the susceptibility at $T\sim 88$~K and
$\sim53$~K, consistent with that reported by Foo
et.al.~\cite{Foo2004}.  Moreover, we find that the anomaly at 88~K
is only apparent in $\chi_{ab}$.

\begin{figure}
\includegraphics[width=3.5in]{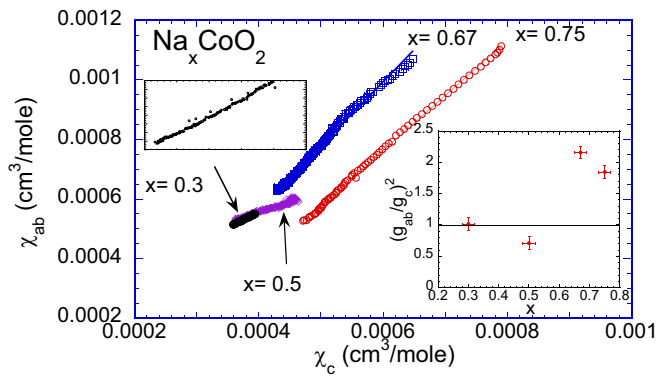}
\caption{\label{fig:eps2} (color online) $\chi_{ab}$ versus
$\chi_{c}$ for Na$_{x}$CoO$_{2}$ with x = 0.75, 0.67, 0.5 and 0.3.
The inset shows the fitted slope which corresponds to
$(g_{ab}/g_{c})^{2}$ as described in the text.}
\end{figure}

These measurements on single crystals yield additional information
on the spin anisotropy for the various compositions.  In
Fig.~\ref{fig:eps2}, plots of $\chi_{ab}$ versus $\chi_{c}$ are
shown for all crystals for temperatures between 50 and 250 K,
where $T$ is an implicit parameter. The $\chi_{ab}$ versus
$\chi_{c}$ curve shows a remarkably linear relationship for all
$x$.  This linear relationship holds both for samples with
Curie-Weiss behavior ($x>0.5$) and for samples with $\chi$
increasing with increasing $T$ ($x<0.5$), albeit with different
magnitudes of the slope.

One way to parameterize this behavior for all of our samples is
with the following analysis.  This analysis allows us to extract
information from the slopes of the curves in Fig.~\ref{fig:eps2},
without assuming a specific temperature-dependence of $\chi$.  The
measured susceptibility is composed of a temperature-independent
contribution $\chi_{\circ}$ (which includes the Van Vleck
paramagnetism and the core diamagnetism) and a
temperature-dependent term $\chi_{e}(T)$ (which represents
contributions from either localized spins or the spin response of
delocalized electrons): $\chi_{ab,c}(T) =
\chi_{\circ}^{ab,c}+\chi_{e}^{ab,c}(T)$.  We then write the
temperature-dependent term in the form $\chi_{e}^{ab,c}(T) =
(g_{ab,c})^{2} f(T)$, which assumes that the anisotropy of the
spin susceptibility $\chi_{e}$ results from an anisotropic
$g$-factor. Note that the $g$-factor for localized spins is
related to the spin-orbit coupling, and the effective $g$-factor
for delocalized electrons is related to the coupling between the
applied field and the total angular momentum of the system.  This
leads to the following relation between $\chi_{ab}(T)$ and
$\chi_{c}(T)$:
\[\chi_{ab}(T) = (g_{ab}/g_{c})^{2} \chi_{c}(T) +
[\chi_{\circ}^{ab} - (g_{ab}/g_{c})^{2} \chi_{\circ}^{c}].\]

\noindent The main point of this analysis is that the fitted slope
of the data in Fig.~\ref{fig:eps2} corresponds to the ratio
$(g_{ab}/g_{c})^{2}$. This ratio is plotted as a function of $x$
in the inset of Fig.~\ref{fig:eps2}. The sample with $x = 0.67$
has the largest anisotropy of $g_{ab}/g_{c}$ $\sim$ 1.45, whereas
the sample with $x = 0.3$ is nearly isotropic.  This behavior
further highlights the unusual metallic state which exists in the
phase diagram for $x>0.5$.

The samples with $x>0.5$ can be further analyzed by fitting the
high-temperature susceptibility to a Curie-Weiss law, $\chi =
\chi_{\circ} + C/T$.  The fits were performed over the range $T =
50 - 250$~K, and the fit parameters for both field orientations,
as well as the powder average, are shown in Table 1. For
comparison purposes, results from data taken on a flux-grown
crystal with $x = 0.7$ is also shown.  We find that the results of
the fits have a small dependence on the choice of the temperature
range selected.  Overall, however, the fits appear to converge
with an error less than 15\%.  The temperature-independent value
for $\chi_{\circ}$ agrees well with the estimate of the orbital
contribution ($\sim$2*10$^{-4}$ cm$^{3}$/mole) from Knight shift
analysis of $^{59}$Co NMR.~\cite{Imai-unpub}

\begin{table}
\caption{\label{tab:table1}Curie-Weiss fitting parameters for Na$_x$CoO$_2$ }
\begin{ruledtabular}
\begin{tabular}{ccccccc}
&&x=0.75&x=0.70\footnotemark[1]&x=0.67\\
\hline
$\parallel$ab & $\chi_{\circ}$ & .000187 & .000268 & .000375\\

$\parallel$ab & C & .181 & .145 & .0677\\
$\parallel$ab & $\theta$ & -130 & -103 & -46.9\\
$\parallel$c & $\chi_{\circ}$ & .000326 & .000223 & .000358\\
$\parallel$c & C & .0751 & .0665 & .0183\\
$\parallel$c & $\theta$ & -93 & -64.5 & -12.0\\
Powder & $\chi_{\circ}$ & .000231 & .000267 & .000356\\
Powder & C & .147 & .112 & .0554\\
Powder & $\theta$ & -125 & -87.8 & -46.9\\

\end{tabular}
\end{ruledtabular}
\footnotetext[1]{flux grown}
\end{table}

\begin{figure}
\includegraphics[width=3.1in]{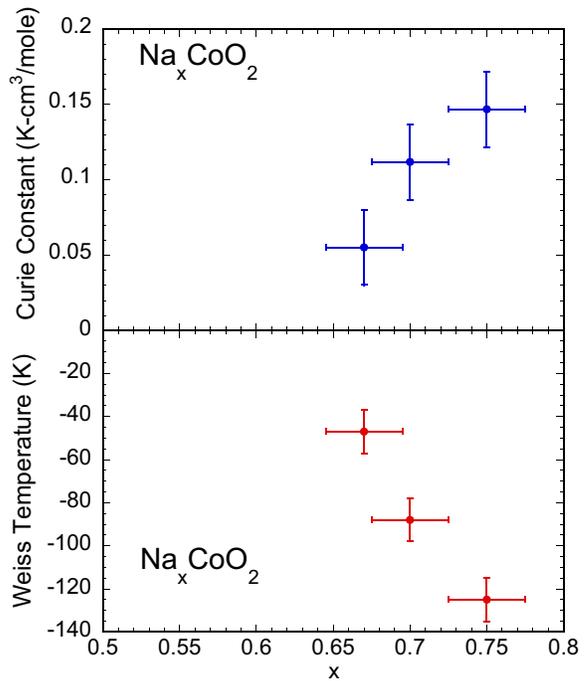}
\caption{\label{fig:eps3} (color online) Top panel: Curie constant
versus x. Bottom panel: the Weiss temperature versus x.  Both are
taken from powder averaged values shown in Table \ref{tab:table1}}
\end{figure}

The fitted values for the powder average of the single crystal
with $x = 0.75$ yield a Curie constant of $C \sim$ 0.147
K-cm$^{3}$/mole, $\chi_{\circ}\sim 2*10^{-4}$ cm$^{3}$/mole, and
Weiss temperature $\theta \sim -125$~K.  We have examined the
validity of these results for the single crystal sample by
measuring five different batches of powder Na$_{0.75}$CoO$_2$
samples (not shown).  All Curie constants fall reliably near
0.149$\pm$0.025 K-cm$^{3}$/mole.  As expected, we find that the
Curie constants for $H\parallel c$ and $H\parallel ab$ are
significantly different for the crystal sample.  This may arise
from an anisotropic $g$-factor as discussed above.  In addition,
it is likely that part of the axial variation of $\chi$ results
from spin anisotropy of the localized Co$^{4+}$ moment.

The effective moment of the Co$^{4+}$ ion can be calculated from
the powder averaged Curie constant. If we assume that the
Curie-Weiss behavior originates from the formal $1-x$ fraction of
Co$^{4+}$ moments with $S=1/2$, then the effective moment
$\mu_{eff}=g\surd{(S(S+1)}\mu_{B} \simeq 2.2~\mu_{B}$ and $g
\simeq 2.5$ for the $x=0.75$ sample.  This value for the powder
averaged $g$-factor, where $g^{2} =
(2/3*g_{ab}^{2}+1/3*g_{c}^{2})$, can be used to estimate $g_c$ and
$g_{ab}$ using our data for $\chi_{ab}(T)/\chi_{c}(T)$.  We obtain
the values $g_{ab}\simeq 2.7$ and $g_{c} \simeq 2.0$.
Alternatively, the value of the Curie constant is consistent with
an interpretation in which the effective local moment is
$\sim$1.1~$\mu_{B}$ averaged over all of the Co ions. In either
interpretation, the coexistence of localized spins and metallic
behavior in this compound remains an intriguing issue to
understand.

The results of our fits to the Curie-Weiss law are summarized in
Fig.~\ref{fig:eps3}. The Curie constant decreases precipitously
with decreasing $x$. This implies that while almost all of the
Co$^{4+}$ spins are localized for $x = 0.75$ (assuming the value
$g \simeq 2.5$ from above), the fraction of localized spins drops
sharply as the Na content is reduced.  Local moment behavior
disappears almost entirely for $x = 0.5$. In parallel with the
loss of local moments, the magnitude of the Weiss temperature
decreases drastically with decreasing $x$. The Weiss temperature
of about -125 K for $x = 0.75$ suggests antiferromagnetic (AF)
correlations between Co$^{4+}$ spins. There is a clear reduction
in the strength of the AF correlations as $x$ decreases towards $x
= 0.5$. These results demonstrate that de-intercalating Na from
Na$_{0.75}$CoO$_2$ modifies the spin system from one described by
localized spins to one described by delocalized spins with weaker
magnetic coupling.

We find the susceptibilities of the two samples with $x=0.3$ and
$x=0.5$ are almost identical (within the errors) above $\sim$100~K
as shown in Fig.~\ref{fig:eps1}, similar to that reported
previously in powder samples.~\cite{Foo2004}  Interestingly, we
find that our non-hydrated Na$_{0.3}$CoO$_2$ sample develops small
anomalies near $T=53$~K and $T=88$~K, after the crystal was stored
in air for more than 6 months.  This suggests that some degree of
Na phase separation may occur over long time scales.  In our
Na$_{0.5}$CoO$_2$ sample, the susceptibility cusp at $T\simeq
53$~K is nearly isotropic, whereas the one near $T\simeq 88$~K is
clearly anisotropic.

\subsubsection{\label{sec:level2}Magnetic susceptibility of
Na$_{0.3}$CoO$_2$$\cdot$yH$_2$O}

\begin{figure}[h]
\includegraphics[width=3.5in]{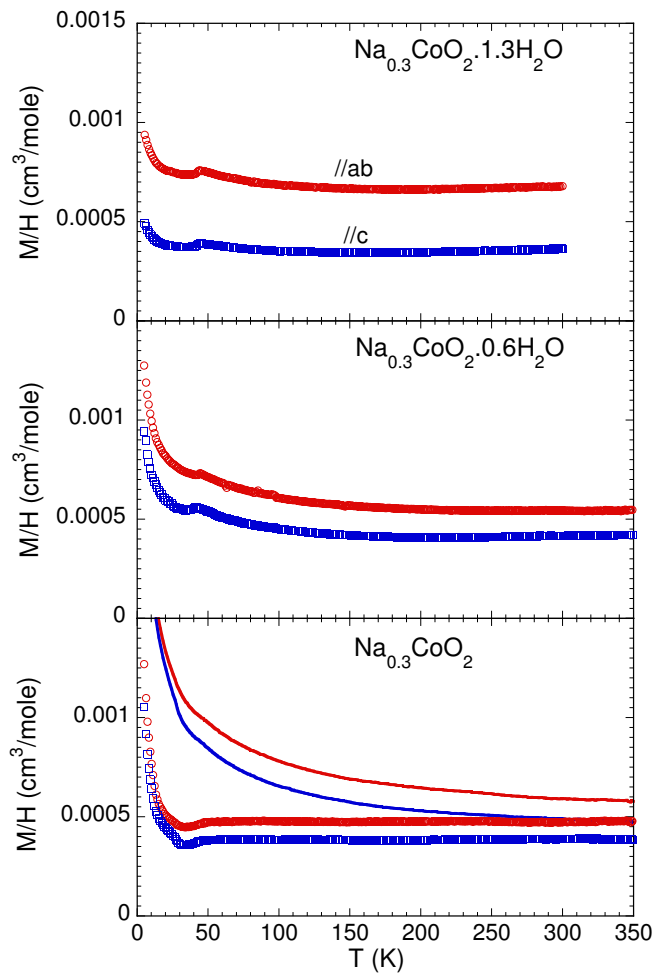}
\caption{\label{fig:eps4} (color online) Magnetic susceptibilities
of Na$_{0.3}$CoO$_2$ $\cdot$yH$_2$O ($y = 0, 0.6,$ and 1.3) under
a magnetic field of 1 Tesla.  The red and blue symbols are for
field applied along the ab and c directions respectively.  The
third panel for Na$_{0.3}$CoO$_2$ shows data both before (lines)
and after (symbols) a Curie tail was subtracted.}
\end{figure}

The magnetic susceptibilities of single crystal samples of
Na$_{0.3}$CoO$_2$$\cdot y$H$_2$O (with $y \simeq 0, 0.6, 1.3$) are
shown in Fig.~\ref{fig:eps4}.  These data were taken on a single
sample which originally had the composition
Na$_{0.3}$CoO$_2$$\cdot$1.3H$_2$O and was subsequently annealed to
reduce the water content to $y=0.6$ and then $y=0$.  All three
samples show nearly temperature-independent susceptibilities with
weak Curie-Weiss-like behavior developing below $\sim 200$~K.   As
water is driven out, the anisotropy of the susceptibility is
reduced, and, at the same time, the Curie behavior above 50~K
becomes slightly enhanced. In addition, the fully dehydrated
crystal is observed to have a small Co$_3$O$_4$ impurity phase,
likely caused by the dehydration process due to partial
decomposition.  By fitting the susceptibility above $T=50$~K for
the fully dehydrated crystal to a Curie law, we find that the
Curie constant is isotropic and corresponds to about $\sim 6$\%\
isolated Co$^{4+}$ ions.  We can therefore identify this as an
impurity contribution, and this term has been subtracted from the
data in the bottom panel of Fig.~\ref{fig:eps4}.  The Curie
corrected susceptibility data for this fully dehydrated
Na$_{0.3}$CoO$_2$ crystal are very similar to the data for the
non-hydrated Na$_{0.3}$CoO$_2$ crystal shown in
Fig.~\ref{fig:eps1} (which do not require this correction).  There
is a cusp in the susceptibility near T $\sim$ 42 K for both the
fully hydrated and partially hydrated crystal, which we will
discuss further below.

\begin{table}
\caption{\label{tab:table2}Curie-Weiss fitting parameters for
Na$_{0.3}$CoO$_2$.$\cdot$yH$_2$O}
\begin{ruledtabular}
\begin{tabular}{ccccc}
&&y=1.3&y=0.6&y=0\\
 \hline
 $\parallel$ab & $\chi_{\circ}$ & .000634 & .000472 &  .000476 \\
 $\parallel$ab & C & .00474 &  .0167 &  .0385\\
$\parallel$ab & $\theta$ & 10.2 &  -20.6 &  -26.7\\
$\parallel$c & $\chi_{\circ}$ & .000334 & .000360 & .000383 \\
$\parallel$c & C & .00160 & .00954 & .0327 \\
$\parallel$c & $\theta$ & 21.2  & -4.12  & -20.4 \\
Powder & $\chi_{\circ}$ & .000533 & .000444  & .000441 \\
Powder & C & .00363  & .0119  & .0380 \\
Powder & $\theta$ & 14.0  & -5.93 & -27.6 \\
\end{tabular}
\end{ruledtabular}
\end{table}

\begin{figure}[h]
\includegraphics[width=3.5in]{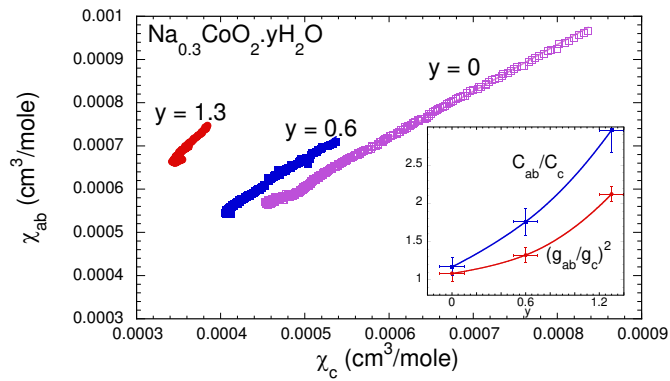}
\caption{\label{fig:eps5} (color online) $\chi_{ab}$ versus
$\chi_{c}$ for Na$_{0.3}$CoO$_2$$\cdot$yH$_2$O with y = 0, 0.6 and
1.3.  The inset shows the fitted slope which corresponds to
$(g_{ab}/g_{c})^{2}$ and Curie constant ratio (C$_{ab}$/C$_{c}$)
as described in the text.  The lines serve as guides for the eye.}
\end{figure}

In Fig.~\ref{fig:eps5}, we plot $\chi_{ab}$ versus $\chi_{c}$,
which again has a remarkably linear dependence.  An anisotropic
Curie-Weiss law $\chi_{\circ} + C/(T-\theta)$ has been used to fit
both directions.  The complete list of fit parameters (above 50K)
is shown in Table 2.  Both $C_{ab}/C_{c}$ and $(g_{ab}/g_{c})^{2}$
are plotted in the inset of Fig.~\ref{fig:eps5}.  We find that the
ratio $(g_{ab}/g_{c})^{2}$ is greater than 1 for both the fully
hydrated and partially hydrated samples. This indicates that the
paramagnetic moments which give rise to the Curie-Weiss behavior
above 50~K are not isotropic, but are strongly affected by the
anisotropic orbital environment.  Hence, this magnetic signal
likely is intrinsic to the crystalline phase.  Such Curie-Weiss
behavior is absent in the non-hydrated Na$_{0.3}$CoO$_2$ sample
shown previously in Fig.~\ref{fig:eps1}. The Curie constant for
the fully hydrated Na$_{0.3}$CoO$_2$$\cdot$1.3H$_2$O crystal
($\sim$ 0.0036 K-cm$^{3}$/mole) suggests about $\sim$1\% of the
spins ($S = 1/2$ and $g^{2} \sim$ 6.3) in the system are
localized. These localized spins may be related to defects formed
during hydration as a result of local structure deformation.
However, their origin remains a topic for further investigation.
Enhanced Curie-like behavior has been reported in powder samples
prepared with the chemical Br$_{2}$ de-intercalation method by
Sakurai and coworkers.~\cite{Sakurai2003}  We note that the Curie
behavior observed below 30~K in our samples has a similar degree
of anisotropy as the high-temperature susceptibility.  Hence, this
behavior likely originates from defects intrinsic to the
crystalline phase.

Two major differences are made apparent by comparing the
susceptibility data of the fully hydrated (FH)
Na$_{0.3}$CoO$_2$$\cdot$1.3H$_2$O and non-hydrated
Na$_{0.3}$CoO$_2$ crystals displayed in Fig.~\ref{fig:eps1} and
Fig.~\ref{fig:eps4}. First, the FH sample has a larger anisotropy.
Second, the FH sample has a larger total susceptibility. The
larger anisotropy likely results from the structural changes of
the CoO layers caused by the hydration.  The higher susceptibility
may result from a change in $\chi_{vv}$ or $\chi_{pauli}$,
although we cannot separate these two contributions independently.
We note that NMR measurements show that the $^{59}$Co Knight shift
is significantly enhanced upon hydration,~\cite{Imai-unpub}
consistent with our findings.  If the susceptibility increase is
dominated by $\chi_{vv}$, a smaller $t_{2g}$ splitting would
indicate a less distorted CoO$_{6}$ octahedra.  However, the
neutron powder results of Lynn {\it{et~ al.}} indicate that a
fully hydrated sample actually has a larger octahedral distortion
compared with a non-hydrated one.~\cite{Lynn2003}  On the other
hand, if $\chi_{pauli}$ is dominant, this implies that the fully
hydrated sample has a higher density of states at the Fermi level.

\section{\label{sec:level1}Discussion and Conclusions\protect\\ }

In the cuprate superconductors, it is generally found that $g_{c}$
is larger than $g_{ab}$.~\cite{Johnston-anisotropic,Watanabe2000}
We find that Na$_{x}$CoO$_{2} \cdot 1.3$H$_2$O has the opposite
relation $(g_{ab}/g_{c})>1$.  A significant difference between
these families of materials is the orientation of the oxygen
octahedra around the Cu and Co ions. In La$_2$CuO$_4$, for
example, the octahedral Cu-O axes are nearly parallel with the
tetragonal crystallographic axes.  On the other hand, in
Na$_{x}$CoO$_{2} \cdot y$H$_2$O, the $z$-axis of the crystal field
(in Co-O coordinates) is tilted away from the crystallographic
$c$-direction by nearly $\sim 60^{\circ}$.  In addition, the
CoO$_6$ octahedra are distorted (compressed along the (111)
direction in Co-O coordinates) from the ideal
configuration.~\cite{Lynn2003,Jorgensen2003}

The $g$-tensor can be calculated from $g_{\mu\nu} = 2
(\delta_{\mu\nu} - \lambda\Lambda_{\mu\nu}$), where $\lambda$ is
the spin-orbit coupling constant and $\Lambda_{\mu\nu} =
\sum_{n}\frac{<0|L_{\mu}|n><n|L_{\nu}|0>}{E_{n}-E_{0}}$.~\cite{White-book}
In a simple ionic picture, each Co$^{4+}$ ion is in the low-spin
state with an unpaired electron in the $a_{1g}$ orbital, higher in
energy than the $E_{g}$ doublet by an amount
$\Delta$.~\cite{QWang2003}  In order to calculate the $g$-tensor
with respect to the crystallographic axes, the orbital wave
functions can be transformed from the original Co-O coordinates
within the CoO$_{6}$ octahedra to the crystal coordinates in the
following way\cite{Zou2004} $|a_{1g}\rangle = d_{3z^{2}-r^{2}}, \;
|E_{1g}\rangle = (\surd2d_{xy}+d_{yz})/\surd3, \;$ and
$|E_{2g}\rangle = (-\surd2d_{x^{2}-y^{2}}+d_{xz})/\surd3$.  Using
this, we find that $g_{c} = 2$, independent of the
$\lambda/\Delta$ ratio.  This is consistent with our estimate of
$g_c$ from the analysis of our susceptibility data.  The values
for $g_{a}$ and $g_{b}$ are expected to be larger than 2 so long
as $\lambda$ is negative.  We note that this model neglects
effects due to hybridization, which may be
significant.~\cite{Marianetti2004}  Recent local density
approximation (LDA) calculations by Marianetti and
coworkers~\cite{Marianetti2004} show that electrons in the
$t_{2g}$ orbitals are accompanied by a redistribution of the
charge in the hybridized $e_g$ and oxygen orbitals.

We remark further on the 42~K anomaly observed in both fully
hydrated Na$_{0.3}$CoO$_2 \cdot 1.3$H$_2$O and partially hydrated
Na$_{0.3}$CoO$_2 \cdot 0.6$H$_2$O crystals (shown previously in
Fig. 4). This anomaly is much weaker or absent in the fully
dehydrated and non-hydrated Na$_{0.3}$CoO$_2$ crystals. Hence, we
conclude that the anomaly results from the hydration process.
However, it is not clear at this point if this reflects the
intrinsic behavior of the hydrated sample, or if it originates
from an impurity phase. We note that the 42~K anomaly does not
exist in fully hydrated powder samples, irrespective of chemical
or electrochemical de-intercalation. It is therefore tempting to
assign the 42~K anomaly to the existence of a Co$_3$O$_4$
impurity, especially since the bulk AF transition temperature of
Co$_3$O$_4$ occurs around $\sim 33$~K.  However, hysteretic
behavior is observed near 42 K (not shown), which is not expected
from a Co$_3$O$_4$ impurity phase. Sasaki and coworkers have also
reported a 42~K anomaly in hydrated single crystal
samples.~\cite{Sasaki2004} They proposed that it originates from
residual oxygen on the surface of the crystal. However, if this is
only a surface effect, powder samples should show a more
pronounced 42~K anomaly under the same treatment. Unlike studies
on K$_x$CoO$_2$ powder samples, where a Co$_3$O$_4$ inclusion can
be reliably subtracted from the total
susceptibility,~\cite{Nakamura1996} we cannot completely eliminate
the anomaly near 42~K with a simple subtraction of a Co$_3$O$_4$
($T_N \sim 33$K) impurity phase.

In summary, we have presented a systematic study of the magnetic
susceptibility of Na$_x$CoO$_2$ $\cdot$yH$_2$O (with 0.3 $<$ x $<$
0.75 and y $\sim$ 0, 0.6 and 1.3) using single crystal samples.
For non-hydrated samples, we find that the derived anisotropic
g-factor ratio $(g_{ab}/g_{c})$ decreases significantly as the
composition is changed from the Curie-Weiss metal with $x = 0.75$
to the paramagnetic metal with $x = 0.3$.  We confirm that a model
of localized Co$^{4+}$ spins describes the spin susceptibility of
Na$_{0.75}$CoO$_2$.  However, the fraction of localized spins
decreases precipitously upon de-intercalation.  For the
composition with x=0.3, the anisotropy in the susceptibility
becomes more pronounced with increasing hydration.  In addition,
the magnitude of the susceptibility is larger in fully hydrated
Na$_{0.3}$CoO$_2$$\cdot$1.3H$_2$O than in non-hydrated
Na$_{0.3}$CoO$_2$.  The hydrated crystals also contain a small
additional fraction of anisotropic localized spins.  These results
provide a new piece to the picture of how the spin behavior
evolves with changing Na content and water content.

\begin{acknowledgments}
We thank P.A. Lee and T. Imai for many insightful discussions.
This work was supported primarily by the MRSEC Program of the
National Science Foundation under award number DMR-02-13282.
Y.S.L. also acknowledges support by the National Science
Foundation under Grant No. DMR 0239377.  J.H.C. was partially
supported by Grant No. (R01-2000-000-00029-0) from the Basic
Research Program of the Korea Science and Engineering Foundation.
\end{acknowledgments}

\bibliography{NaxCoO2}

\end{document}